\documentclass[10pt, aps,prd, nofootinbib, preprintnumbers,
  twocolumn, superscriptaddress]{revtex4-2}
\pdfoutput=1

\usepackage[pdftex]{graphicx}
\usepackage{amssymb, amsmath, bm}
\usepackage{microtype}
\usepackage{etoolbox}
\apptocmd{\sloppy}{\hbadness 10000\relax}{}{}

\usepackage[colorlinks,linktocpage,hypertexnames=false]{hyperref}
\hypersetup{colorlinks=true, citecolor=blue}

\begin{document}

\title{New Method for Investigating the Presence of Extragalactic Magnetic Fields}

\author{ B. Stern}\email{stern@inr.ru}, 
\affiliation{Institute for Nuclear Research of the Russian Academy of Sciences, Moscow 117312, Russia}
\author{I. Tkachev}\email{tkachev@inr.ru}
\affiliation{Institute for Nuclear Research of the Russian Academy of Sciences, Moscow 117312, Russia}
\affiliation{Physics Department and Laboratory of Cosmology and Elementary Particle Physics, Novosibirsk State University, Novosibirsk 630090, Russia}

\begin{abstract}
The extragalactic magnetic field could be detected by searching for signatures of the electromagnetic cascade initiated by high-energy photons on the intergalactic radiation and deflected by the field. This process produces a time delay and an extended gamma-ray halo around the source, which are looked for. We propose a new signature of electromagnetic echoes: the asymmetry of the gamma-ray distribution around blazars. As a measure of asymmetry, we use the offset of the gamma-ray distribution to the location of the blazar. This offset is due to the tilt of the jet of the blazar relative to the line of sight. Using a subsample of the 10 brightest BL Lacs, we exclude  the range of extragalactic magnetic fields from  $10^{-16}$ to $10^{-14}$ G, assuming that these objects have maintained a constant average luminosity over hundreds of thousands of years.
 \end{abstract}

\maketitle


\section{Introduction}

There is no theoretical consensus on the origin and likely values of the extragalactic magnetic field (EGMF), for an overview see~\cite{durrer2013cosmological}. There is a well-established upper limit on its value of $\sim10^{-9}$ G, obtained from Faraday rotation measurements (see, e.g.,~\cite{PhysRevLett.116.191302}), and a number of papers proclaiming lower limits.

Two approaches are used to test EGMF and establish lower limits. Both are associated with the $\gamma$-ray echo caused by electromagnetic cascades  initiated by multi-TeV photons on extragalactic background light. The first one is to search for delayed photons after gamma-ray flares of blazars or GRBs. This approach was suggested by ~\cite{Plaga} and developed by ~\cite{PhysRevD.80.123012, Dermer_2011}. The lower limit set by this method is of the order of $10^{-17}$ G~\cite{Ichiki_2008, Dzhatdoev:2023opo, Acciari_2023, 2024A&A...683A..25V}.
The second approach is to search for extended halos around the blazar's gamma-ray images due to angular spread caused by the same process~\cite{10.1093/mnras/264.1.191,1994ApJ...423L...5A}. Most studies of this effect report the absence of angular spreading and put a lower limit on the EGMF value at the level $10^{-15}$ G (\cite{doi:10.1126/science.1184192, 10.1093/mnrasl/slad142, Vovk_2012, Ackermann_2018, HESS:2023zwb}).   We stress that these bounds depend on the $\gamma$-ray energy and EGMF correlation length, see e.g. \cite{Acciari_2023}. Comprehensive reviews with theoretical background can be found in~\cite{Neronov:2009gh,universe7070223}. 

 There are several reports of observations of extended gamma-ray halos around BL Lacs, for example, a signal in stacked images of a sample of active galactic nuclei was detected in~\cite{Ando_2010}. However, Ref.~\cite{refId0} demonstrated, and \cite{Fermi-LAT:2013mql} confirmed, that this was merely an underestimation of the telescope's point spread function (PSF) width. Then, in the paper~\cite{PhysRevLett.115.211103}  an interesting hint  of  image broadening was found in a stacked sample of high synchrotron-peaked BL Lacs.

This paper presents a new method for detecting the presence of EGMF. It is based on the asymmetry of the blazar image rather than its extent, and is sensitive enough to give a significant signal from an individual object.  We limit the present study to  {\it Fermi} LAT data and one simple characterization of image asymmetry.

The paper is organized as follows. In Section~\ref{sec:model} we formulate the problem and describe the Monte-Carlo model. In Section~\ref{sec:model_results} we present the results of numerical simulations and describe the expected signal for different model parameters. 
In Section~\ref{sec:constraints}, we describe how we formed our sample of BL Lac objects and discuss the challenge of non-unique definition of their locations. Further in this section we put the constraints on the extragalactic magnetic field using the brightest part of the sample. In Section~\ref{sec:discussion}  we discuss the advantages of the method and perspectives of further efforts in this direction. Additional technical information is provided in the Appendix.

\section{The method  and the model}
\label{sec:model}

Astrophysical high-energy photons are converted into $e^+e^-$ pairs by interacting with the extragalactic background light, which is optically thick for photons in the TeV range at distances of hundreds of Mpc. In modelling this process, we use an estimate of the spectral density of infrared-optical photons in intergalactic space from~\cite{refId0R}. The produced pairs subsequently comptonize CMB photons up to the GeV energy range covered by the gamma-ray telescopes. The path length of an electron before it can Comptonize the relic radiation  is about 10 kpc, which is negligible compared to the distance to the source. This cooling distance limits the range of magnetic field values that can be detected using this technique to $10^{-16} - 10^{-14}$~G. A weaker field is unable to deflect the electrons by an angle sufficient to distort the blazar's image;  a stronger field will deflect them so that the Comptonized photons will lose their connection with the source and contribute to the isotropic gamma-ray background. The process in this range  extends a blazar image, producing a halo. 

\begin{figure}
	\includegraphics[width=\columnwidth]{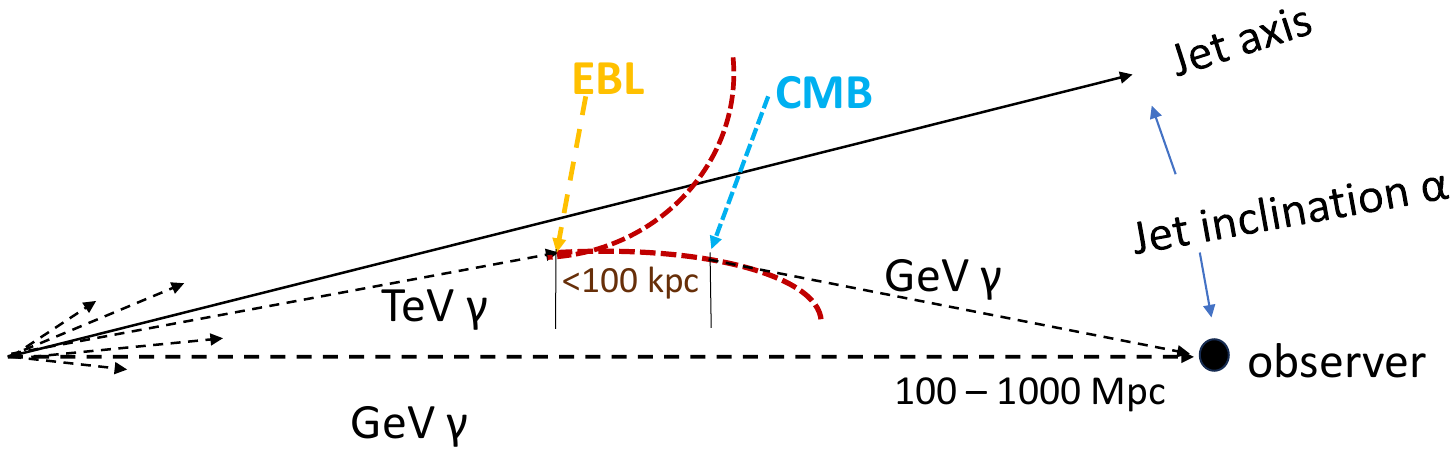}
    \caption{
Diagram illustrating the formation of a "cascade echo". The primary TeV radiation in the jet creates $e^+e^-$ pairs in the intergalactic background light; $e^+e^-$ pairs deflected by EGMF create secondary GeV photons when interacting with the cosmic microwave background radiation.
The resulting GeV halo is shifted towards the jet inclination. 
}
\label{fig:scheme}
\end{figure}

The central point of the new method we propose is that the halo must be asymmetric, since the blazar jets are not directed precisely at us, see Fig.~\ref{fig:scheme}. 
We suggest that halo asymmetry may be a more efficient indicator of the intergalactic field than the image extent above the PSF.
In particular, the new approach does not rely on the perhaps poorly known energy-dependent PSF. 

We derive the expected properties of the asymmetric halo using Monte Carlo simulations. To do this, we first need a blazar model. We make the following assumptions:
\begin{itemize}
  \item[i)] The blazar jet has a Lorentz factor of 20. This is a typical value estimated from the superluminal motion (\cite{Jorstad_2001}).
  \item[ii)] The jet is strongly collimated. The angular distribution of photons results from the isotropic motion of emitted particles in the comoving reference frame of the jet.
 \item[iii)] The spectral energy distribution is a flat power law, i.e, the photon number index is $\Gamma = -2$. This is a typical feature of BL Lacs. Some objects (e.g. Mkn 421) have harder spectra in the 1 GeV - 1 TeV range, while others (e.g. BL Lac) are softer. Flat spectrum is a good compromise.
 \item[iv)]  The maximum photon energy is 10 TeV in the jet comoving frame. 
 \item[v)] We model the EGMF in the following way. From birth until the loss (due to cooling) of the ability to produce Compton photons with energy greater than GeV, each electron moves in a field with an initially randomly oriented but then constant transverse component H with a given amplitude. This modeling scheme corresponds to a correlation length   $\gtrsim $ Mpc.
A shorter correlation length will reduce the effect we are looking for if the field is in the range of $10^{-16} - 10^{-15}$ G, and will enhance it in a stronger field.    
\end{itemize}
Items iii and iv imply low external radiation density at the source, this condition is natural for BL-Lacs. 
This basic model does not cover the entire variety of BL-Lacs and EGMF, but it is quite sufficient for studying the proposed method for detecting extragalactic magnetic fields. Therefore, we do not change the initial spectrum of photons or their angular distribution. We only change the distance to the source and the orientation of the jet relative to the line of sight (inclination angle). 

We simulate the photon interactions and their propagation from the source using a direct Monte Carlo method without statistical weights, applying fragments of the code described in ~\cite{10.1093/mnras/272.2.291}.
The energy, direction and delay of each arriving photon are recorded on a "detector". The angular location of each incoming photon relative to the line of sight was then randomized using a template PSF.  The latter was constructed using a real photon image of Mkn 421 with energies above 1 GeV, consisting of $\sim$ 19,000 photons. This template image was cropped at 1.5$^\circ$ from the center with background subtraction.  

\section{The results of Monte-Carlo simulations}
\label{sec:model_results}

In modeling, we can distinguish between direct photons which come from a source without interaction, and echo photons, which come as a result of cascade deflection in the EGMF see Fig.~\ref{fig:scheme}.  Clearly, the echo is asymmetric in position angle, being shifted towards the jet inclination.
Echo photons also arrive with a time delay up to $10^5$ - $10^6$ years. For variable sources this can be very important when searching for the EGMF signature. 

\begin{figure}
		\includegraphics[width=\columnwidth]{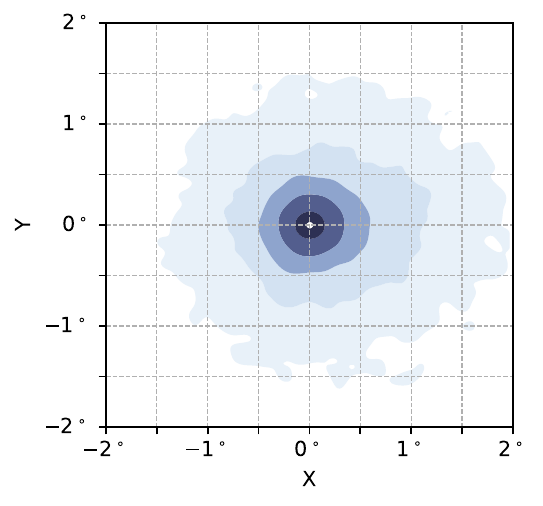}
 \caption{
A simulated color-coded map of counts in the plain orthogonal to the line of sight for the case z = 0.1, H=10$^{-15}$ G, jet inclination  $\alpha = 3^\circ$ directed to the right.}
\label{fig:map_simul}
\end{figure}

In real astrophysical data, direct and echo gammas  are indistinguishable. However, their sum, which is observable, will be statistically different to the left and right of the source position, see Fig.~\ref{fig:map_simul}. 

We found that Kolmogorov-Smirnov type tests for this bivariate distribution are a very sensitive tool for assessing skewness.This imposes strong restrictions on EGMF even for individual BLLs and may serve as a basis for its discovery. The results of our in-depth research on specific BLLs will be presented in our upcoming paper~\cite{StTk}.

In this paper, we apply a simpler alternative measure of the significance of asymmetry based on the offset  of the blazar halo relative to the known location of the source. In a given fixed field of view around a source, the offset O is defined as the average value of the photon coordinates   
\begin{equation}
O^2 \equiv X^2 + Y^2,
\label{eq:O}
\end{equation}
where $X = \Sigma_{ i}^N  x_i /N$ and $Y = \Sigma_{ i}^N  y_i /N$.  
Variables $X$ and $Y$ are normally distributed with a standard deviation $\sigma/\sqrt{N}$ if $x_i$ and $y_i$ are distributed with  $\sigma$. Then, $O^2$ follows a $\chi^2$-distribution, while $O$ is Rayleigh-distributed. 
Despite weaker statistical power for an individual source compared to KS-tests~\cite{StTk},   statistics based on the $O^2$  variable have the advantage of allowing straightforward   stacking  of multiple BLLs, ensuring the meaningful significance of the EGMF constraints.

On the other hand, the statistics based on offsets is prone to background inhomogeneities and possible systematics in data that mimic EGMF. These issues are extremely important when indications of EGMF are found. As we will see, we do not find hints of EGMF and instead derive meaningful constraints, which in itself tells us that these interfering effects are small. However, we will discuss them in detail later and in the Appendix~\ref{sec:app_moc}.

The offset, according to our simulations, increases linearly with the field of view  below $1^\circ$ and more slowly above $1^\circ$. On the other hand,  the problems with the background increase quadratically, including the number of interfering sources within and around the field of view. The density of Ferim sources is ~0.18 per square degree and the average distance to the nearest source is $\sim 1.3^\circ$.  As a result of the trade-off between the magnitude of the effect and the level of contamination, we chose a fiducial field of view (FoV) of radius $1^\circ$. We did not change this value to avoid the "look elsewhere" effect.

According to \cite{universe7070223}, the effectiveness of halo detection depends on two factors related to the chosen field of view. First, the size of the extended emission should be such that it is fully contained within FoV. Second, this extension must exceed the angular resolution of the detector, $\theta_{\rm psf}$. In our case, $\theta_{\rm psf} \approx 0.25^\circ$, which can be seen from Fig.~\ref{fig:map_simul}. Inside this region, direct photons dominate, and their distribution is symmetric. We also see that the echo extends to two degrees, but cutting  the FoV to $1^\circ$ does not significantly reduce the collected asymmetry statistics.

\begin{figure}
\includegraphics[width=\columnwidth]{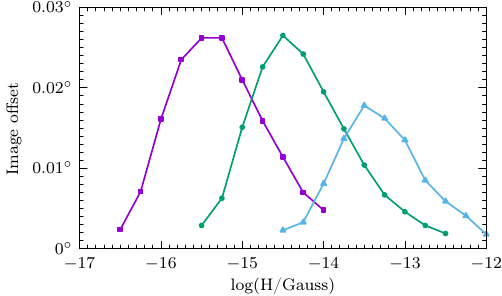}
\caption{
Dependence of the image offset on  the transverse component of the magnetic field for three different  energy thresholds, 1 GeV, 10 GeV, and 100 GeV. The redshift is $z = 0.2$.  
}
\label{fig:offset-H}
\end{figure}

\begin{figure}
\includegraphics[width=\columnwidth]{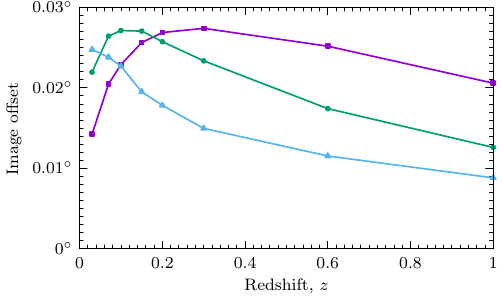}
\caption{
The offset of the source image versus redshift for different energy cuts and magnetic field values. 
Magenta: E $>$ 1 GeV, H $= 3 \times 10^{-16}$ G, green E $>$ 10 GeV, H $= 3 \times10^{-15}$ G, blue: E $>$ 100 GeV, H = $3 \times 10^{-14}$  G. The jet inclination angle is $2^\circ$. }
\label{fig:offset-z} 
\end{figure}

Figures~\ref{fig:offset-H} and~\ref{fig:offset-z}  show the  simulated offset of {\it Fermi} LAT images of BL Lacs for different model parameters.
The dependence of the offset on the magnitude of the transverse magnetic field is shown in Fig.~\ref{fig:offset-H} for different low energy cuts in photon dataset. The range of magnetic fields over which the effect can be detected using {\it Fermi} data is only 2–2.5 orders of magnitude wide.  The Figure~\ref{fig:offset-z}  shows the offset dependence on the redshift of a source for several values of low energy cuts and EGMF magnitude.

As can be seen, the typical offset ($\sim$0.02$^\circ$ in regions where the influence of the EGMF is considerable) is significantly less than the half-width at half-maximum of the PSF  $\sim$0.15$^\circ$. Nevertheless, it is still confidently detectable for bright sources.  In particular, the statistical error in the offset value is $\sim$0.0025$^\circ$ for the brightest sources such as  Mkn 421, and $\sim$0.007$^\circ$ for sources an order of magnitude fainter.

In Figs.~\ref{fig:offset-H} and  \ref{fig:offset-z} we presented the offset assuming that the source is constant during hundreds thousand of years.  If we observe the source in a relatively high state the offset will be smaller, e.g. if it is currently an order of magnitude brighter then  the effect of the extragalactic field will be undetectable for this source. On the other hand, if the source is in a very low state, its GeV photon image could be dominated be the echo and the offset in this case is an order of magnitude larger.

 \section{Constraints on the extragalactic magnetic fields}
\label{sec:constraints}

\paragraph{Data.}

We restricted our analysis to BL Lacs from the {\it Fermi} 4FGL catalogue (1407 objects), which are located in low-background regions (755 objects). We use all photons of class {\it source} above 1 GeV regardless of their conversion type (front and back). The number of such photons is of order of $4 \times10^7$ the entire {\t Fermi} data set.  The background was measured in a ring $1.5^\circ < \theta < 2.5^\circ$ around the object, the cut resulted in 4500 photons above 1 GeV during the 833 weeks of observations that were covered by the data we used. This corresponds to $7.1\times 10^{-6}$ photons per square degree per second. We then apply the following filter: no other 4FGL sources in the $1^\circ$ neighbourhood around the source, and no sources in the ring $1^\circ < \theta < 1.5^\circ$ brighter than 0.1 times the central source's brightness. This filter reduced the sample to 310 objects.

An important question is what should be used as the real source location, since {\it Fermi} and optical localizations are somewhat different. 
The choice of wrong source positions can create artificial asymmetry, reducing or eliminating constraints on EGMF. This can even lead to false signals that require deeper investigation.
Fortunately, we know the true source coordinates with high precision. They are provided by optical or radio observations. However, incorrect asymmetry signals may occur even if one uses optical localization, for example, if the data contains systematics amounting to small global shifts in the derived coordinates. Such systematics do exists in the data, seeAppendix for details. Such systematics will be automatically compensated for by the choice of {\t Fermi} localization. Therefore, first, we conclude that the offsets relative to {\it Fermi}  localizations are expected to be smaller, and the choice of optical localization provides robust conservative constraints on the EGMF. Secondly, it is important to see what happens with both possible choices.

It should be emphasized that the offset, Eq.~\ref{eq:O},  in the presence of EGMF cannot be compensated by the coice of  {\it Fermi} source localization, but should be considered as one of the possible measures of the halo asymmetry, and can be used with either choice of localization.

\begin{figure}
	\includegraphics[width=\columnwidth]{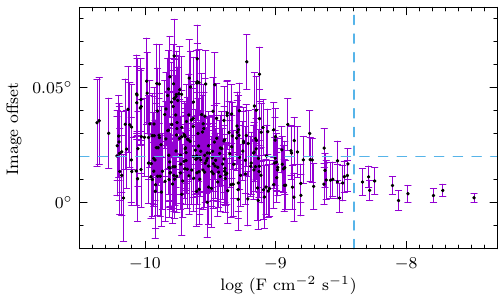}
\caption{
Offsets for 310 BL Lacs selected by low background in the  field of view $\theta = 1^\circ$ versus source brightness. The horizontal dashed line at $0.02^\circ$ indicates a level typical for EGMF-induced  offsets for magnetic field in the range $10^{-16} \lesssim H  \lesssim 10^{-14}$, see Figs.~\ref{fig:offset-H} ,  \ref{fig:offset-z}. 
}
\label{fig:offsets}
\end{figure}
 
The offsets relative {\it Fermi} source  locations for 310 selected BLL as a function of their brightness F (photons cm$^{-2}$s$^{-1}$) are shown in Fig.~\ref{fig:offsets}.  The F values are taken from the 4FGL~\cite{Fermi-LAT:2022byn} catalog. The statistical errors shown in the figure are estimated by randomizing the azimuthal angles of photons in the fiducial field of view. We measure the standard deviation $\sigma$ for the $X$, $Y$ variables, Eq.~\ref{eq:O}, in each object, and then display them as error bars, since they are used in the $\chi^2$ tests later.
 
\paragraph{Constraints on EGMF.}

For the dim end of the BLL distribution, statistical fluctuations and the contribution of real background inhomogeneities are expected to exceed the searched-for EGMF effect. Fig.~\ref{fig:offsets} confirms this.  Therefore, when deriving constraints on magnetic fields, we limited our analysis to the 10 brightest objects. These BLLs are located to the right of the dashed vertical line, their list is given in Table~1. 
\begin{table}
\centering
\small
\begin{tabular}{lllll} 
\hline   
object& z  & flux\\
\hline   
 Mkn 421       & 0.03&3.31\\
 PKS 2155-304 &  0.116&1.91\\
 PKS 0537-441 &  0.894&1.62\\
 PKS 1424+240 &  0.160&1.03\\
 S2 0109+22  &   0.265&0.875\\
1H 1013+498 &   0.212&0.786\\
PKS 2233-148 &  0.33&0.574\\
 PG 1246+586 &    0.847&0.526\\
GB6 J1542+6129&  0.507&0.516\\
PKS 0301-243 &  0.260&0.463\\
\hline
\end{tabular}
\caption{ The sample of brightest BL Lacs used for setting the constraints on magnetic field. The brightness is given in units 10$^{-8}$ photons cm$^{-2}$ s$^{-1}$
according to {\it Fermi} Source Catalogue~\cite{Fermi-LAT:2022byn}.}
\label{tbl:1}
\end{table}

Though Fig.~\ref{fig:offsets} hints by itself at conclusions and BLL selection criteria, we verified in MC simulations that background and limited statistics effects are only important to the left of the vertical line in Fig.~\ref{fig:offsets}. In particular, in the Supplementary Material, the entire BLL sample from Fig.~\ref{fig:offsets} is examined for consistency with the hypothesis of no EGMF effects using implant mock-ups.

We estimate the constraint on the magnetic field in the following way. 
First, we calculate total $\chi_0^2$ for this sample against null hypothesis (no EGMF, i.e. no physical offsets). In calculating $\chi^2$ below, the statistical errors  are estimated by randomizing the azimuthal angles of photons in the fiducial field of view. 

We have two possible null hypothesis options, corresponding to two possible object location procedures.
Total $\chi^2$ for the sample assuming {\t Fermi} locations is $ \chi_0^2=23.45$ for 20 degrees of freedom (two coordinates for each object). If we use optical locations total $\chi_0^2$ is 32.2.  This confirms, in particular, that background inhomogeneities do not significantly contribute to offsets of the 10 brightest BLs.
 
Second, we calculate the expected offsets in the presence of EGMF  for each of 10 blazars of our sample using results of our Monte Carlo simulations described in Section~\ref{sec:model_results}.  We took into account the dependence of the offset on z (taking the real redshift of each object) and on the jet inclination angle, integrating over the latter from 0 to 2$^\circ$.  Then we calculate expected $\chi^2$ for the sample with obtained Monte Carlo offsets using rms error for each object as described above.
 
 The likelihood of a given value of EGMF, common for 10 objects of the sample,  is given by the probability to get $\chi^2 < \chi_0^2$. We use two different energy cutoffs 1 GeV and 10 GeV. Since the subset of photons above the 10 GeV threshold is a small fraction of the 1 GeV fluence, we can treat the corresponding two probabilities as independent factors. 

 \begin{figure}
	\includegraphics[width=\columnwidth]{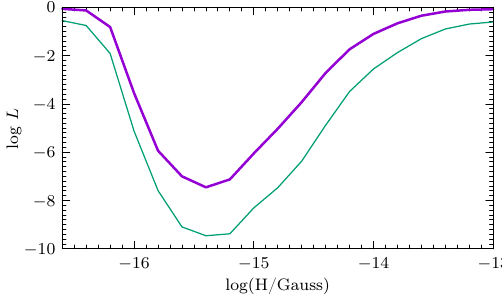}
\caption{ 
The likelihood profile for the extragalactic magnetic field based on the image offsets of the 10 brightest BLL. The thick line represents constrains with the optical source localization, while the thin line represents the {\it Fermi} localization.  
}
\label{fig:rej} 
\end{figure}

The resulting dependence of the likelihood on the magnetic field strength H is shown in Fig. ~\ref{fig:rej}.  
 Note that the right slope of the likelihood profiles is flatter than the left slope. This is due to the contribution of the 10 GeV subsample, which is significant for $10^{-15} \lesssim H  \lesssim 10^{-14}$.  We also estimated the constraint using the 16 brightest objects and got almost the same result.

The difference in EGMF constraints when using optical and {\t Fermi} locations is mainly due to a single source, Mkn~421.  If we exclude it,  then $\chi_0^2$ for the sample assuming {\t Fermi} locations is $ \chi_0^2=22$ for 18 degrees of freedom, whereas if we use optical locations $\chi_0^2$ is 24.2 meaning that the {\it Fermi} systematics discussed in Appendix \ref{sec:app_bias} is not significant for us.  However, inspection of the photon distribution around Mkn~421 reveals significant irregularities in its center~\cite{StTk}. This can most likely be explained by the presence of an unidentified interfering source. However, at present we cannot exclude a hint of EGMF in the direction of Mkn~421. 

In principle, this result could be affected by a brightness selection bias if the gamma-ray luminosity of BL Lacs varies strongly on large time scales because the echo arrives with a time delay.  This ambiguity is inherent to any method of EGMF constraints discussed in the Introduction and is usually not considered.  Note also that the brightest blazar objects in our sample, with the exception of PKS 0537-441, are also the closest ones, so the impact of variability on the selection bias is unlikely to be significant.

\section{Discussion and conclusions}
\label{sec:discussion}

In this paper, we propose a new method for detecting or constraining the EGMF.  Existing methods look  for an excess of events above the telescope's point spread function. Instead, we suggest to search for the asymmetry of the gamma-ray distributions around blazars. The new method does not rely on the energy-dependent PSF's knowledge, which is a clear advantage. 

Various statistical tests can be used to detect asymmetry in bivariate data. In this paper, we have used a simple approach based on the offset of the source image relative to its known astrophysical position. Several sources can be stacked in this approach to improve sensitivity and we exclude  EGMF in the range  from $10^{-16}$ to $10^{-14}$  G at the two sigma level using the ten brightest BLLs. This limit is comparable to that established by HESS and MAGIC Cherenkov telescopes, as reported in \cite{HESS:2023zwb,refId0}. Here, we obtain this constraint using only {\t Fermi} data, and we suggest that our approach could probe a wider range of magnetic fields using Cherenkov arrays. 
 According to the constrained modelling of large scale structure formation and the associated growth  of seed magnetic fields,  EGMF in voids are expected to be on the order of $10^{-13}$~G~\cite{Dolag:2004kp}.  Probing this range requires photons with $E > 100$ GeV, see Fig.~\ref{fig:offset-H}, and Cherenkov arrays are up to the task. 

About the prospects of the new method. Our initial analysis suggests that the statistical tools based on the Kolmogorov-Smirnov test are significantly more sensitive \cite{StTk}. This allows us to derive tight constraints on EMF even for individual bright objects, while simultaneously controlling for interfering nearby sources. If the effect is ever detected, the method can estimate the EGMF along individual paths and measure the inclination angles of the jets.  In reality, blazars can vary on long time scales. This opens the possibility of studying individual dim objects that were bright in the past, during the formation of the gamma-ray echo.
 
 
\begin{acknowledgments}
 The work of  I.T. was supported by the Russian Science Foundation grant 23-42-00066.
\end{acknowledgments}


\appendix

\section{Background  inhomogeneities: investigation using mock blazars}
\label{sec:app_moc}
 
At the low brightness end of the distribution Fig.~5, measured offsets exceed the expected EGMF effect.  In this section, using mock blazars (implants), we show that this is due to the influence of background inhomogeneities.

 At the lowest brightness level, even the limited statistics dominate, therefore,
we cut out the dimest part of the distribution at Log (F) = -9.5, the remaining sample includes 149 objects. The resulting $\chi^2$ for this sample against null hypothesis (no EGMF effect and Gaussian homogeneous background) is 499 per  298 degrees of freedom (two degrees per each object for two coordinates). This means that the null hypothesis can be rejected with high significance ($p < 10^{-6}$) and the question arises, what is the reason for the high $\chi^2$: the echo effect or the background inhomogeneities?

There is a number of significant ($p < 10^{-5})$ offsets for individual objects if we use statistical RMS deviation estimated with randomization of azimuthal angle of photons around the source location. However, visual inspection shows that some of the outliers can be explained by interfering sources  that passed through our filter described in the main paper. 
Can the ``bad'' background alone explain the high $\chi^2$ value, or should we assume that this result is partly due to the  cascade echo effect?         

\begin{figure}[b]
	\includegraphics[width=\columnwidth]{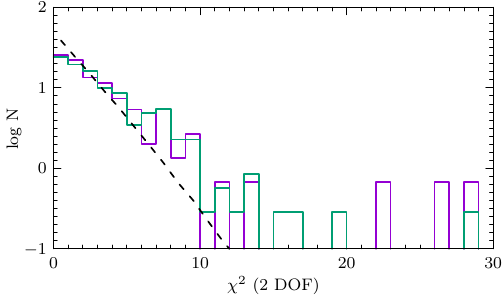}
    \caption{ 
Distributions of  $ \chi^2 $ values for real 149 BL Lacs (magneta histogram) and 350 implants (green histogram). The Gaussian $\chi^2 $ distribution is shown by the dashed line. All distributions are equally normalized.
}
\label{fig:chi2} 
\end{figure}

Although we have cleared the sample of objects with interfering sources from the 4FGL catalog and avoided areas with high background, there may still be unidentified faint sources and a small diffuse background gradient remaining.
Probably, the only way to estimate the effect of background is to plant randomly to the data simulated photons from mock blazars. As a template, we choose 3C 454.3 for its brightness. Then we randomly selected photons from the template with their deviation from the centre and random azimuthal angle. The brightness distribution of implants was the same as of our sample of 149 real objects.   
We set 2000 implants isotropically, 730 of them passed the same low background cut that we used for real BL Lacs. After applying the same nearby source filter as for real BL Lacs, the implant sample was reduced to 350. Fig.~\ref {fig:chi2} shows the distribution of offsets in units of $\chi^2$ for individual BL-Lacs and implants, i.e. the value ${\Delta_x^2}/{\sigma^2} +  {\Delta_y^2}/ {\sigma^2} $ where $\Delta_{x,y}$ is the offset at each coordinate and $\sigma$ is the statistical root-mean-square deviation.    

 One can see, that $\chi^2$ distributions for real BL Lacs and implants are statistically identical. This means that the excessive offsets of GeV images of BL Lacs respectively their optical locations can be entirely prescribed to the inhomogeneous background and a possible contribution of the effect of GeV echo is statistically insignificant.

\section{Bias in  4FGL catalogue versus optical coordinates}
\label{sec:app_bias}

\begin{figure}
	\includegraphics[width=\columnwidth]{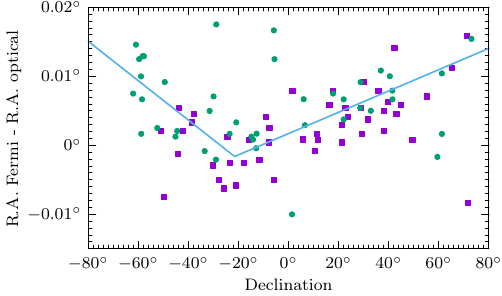 }
\caption{
 Difference in right ascension for sources in 4FGL catalogue and their known optical coordinates versus declination for 50 brightest extragalactic (squares) and 50 brightest galactic (circles) objects. The broken linear function shows the bias estimate. There is no systematic bias in declination. 
}
\label{fig:bias}
\end{figure}

We have found that there is a systematic bias in the locations of {\it Fermi} catalog objects depending on their optical locations, see~Fig.~\ref{fig:bias}. Fermi right ascension is systematically higher than that for optical locations for both galactic and extragalactic objects.  The difference is up to $0.01^\circ$ at a high declination. It appears that this is an overall bias that affects the location of each photon. This offset is much smaller than the angular resolution of the LAT, but is critical to our analysis because it is of the same order of magnitude as the offset we are looking for. The bias mimics an EGMF effect, as if most BL Lac jets are inclined in the same direction.  This is unphysical. For this reason, we adjust the coordinates of each photon to account for this bias, estimating it as a broken linear function  shown in Fig.~\ref{fig:bias}. The resulting constraint on the EGMF with optical source localization, see Fig.~6 of the main paper, moves it closer to the results obtained using {\it Fermi} source coordinates. 

Therefore, the difference between the {\it Fermi} and optical source locations, excluding statistical uncertainties, can be partially attributed to this bias.

\section{Apparently asymmetrical  $\gamma$-sources}

The absence of statistically significant signal of GeV cascade echo in the subsample of bright objects does not mean that the echo cannot be revealed in some individual weaker blazar images.  At the low-luminosity end, disregarding the background contribution, the asymmetry can be large if the source is currently in a low-luminosity state and its image is formed predominantly by echoes of its past bright state.

\begin{figure}
   \includegraphics[width=\columnwidth]{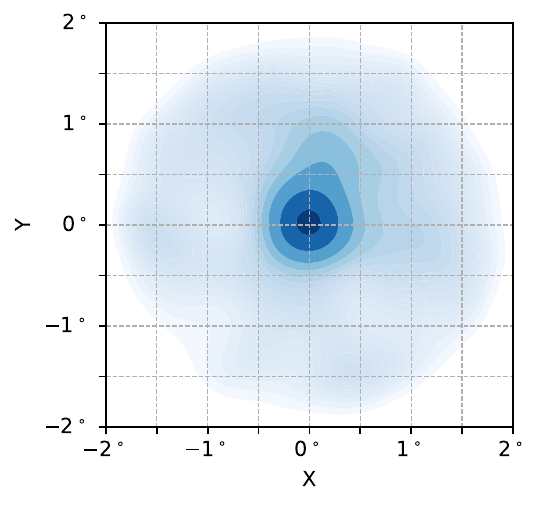}
   \caption{ Count map of photons for PKS 0336-177.  }
\label{fig:PKS}
\end{figure}

We visually inspected BL Lacs images with large, significant offsets, looking for objects with an extended, asymmetric halo that cannot be easily explained by discrete background sources or a background gradient. The relatively weak PKS 0336-177 is an example of this.  The image of this source, Fig.~\ref{fig:PKS}, resembles a jet. However, the projected distance would be  $\sim$ 5 Mpc which seems too large for a MHD jet. In principle, this can be an effect of a gamma-ray beam from the blazar: photons convert not far from the source, and we observe cascade photons. Objects like this require further investigation.

\bibliographystyle{apsrev4-1}
\bibliography{imf}

\end{document}